\documentstyle[12pt]{article} 
\begin{document} 
\begin{titlepage} 
\begin{flushright} 
TPI-MINN-00/21  \\ 
UMN-TH-1904  \\ 
hep-th/0005119
\end{flushright} 
 
\vspace{0.6cm} 
 
\begin{center} 
\Large{{\bf Abrikosov String in {\cal N}=2 Supersymmetric  QED}} 

\vspace{1cm}

Xinrui  Hou
\end{center} 
\vspace{0.3cm}

\begin{center}
{\em Theoretical Physics Institute \\
University of Minnesota\\
Minneapolis, MN 55455, USA}
\end{center} 

\vspace{1cm} 
 
\begin{abstract} 
 
We study the Abrikosov-Nielsen-Olesen string in ${\cal N}=2$ supersymmetric QED 
with ${\cal N}=2$-preserving superpotential, in which case the Abrikosov string 
is found to be $1/2$-BPS saturated. Adding a quadratic small perturbation in the superpotential 
breaks ${\cal N}=2$ supersymmetry to ${\cal N}=1$ supersymmetry. Then the Abrikosov 
string is no longer BPS saturated. The difference between the string tensions for the 
non-BPS and BPS saturated situation is found to be negative to the first order of the 
perturbation parameter. 
\end{abstract} 
\end{titlepage} 
 
\section{Introduction} 
\label{sec1} 
 
The simplest theory where saturated strings exist in the weak coupling regime is 
supersymmetric electrodynamics with the Fayet-Iliopoulos term ~\cite{gm}. 
Topologically stable solutions in this model and its modifications were 
considered more than once in the past ~\cite{gm,c2,c3,c4}. If we consider 
the ${\cal N}=2$ supersymmetric Yang-Mills theory softly broken near the 
monopole or dyon singularities, the Abrikosov strings develop ~\cite{c7}. They were 
discussed in the literature ~\cite{c5,c6,c8,c9} previously. We can see them in 
the effective Lagrangians near the singularities ~\cite{c3}, where the 
superpotential for the monoploes (or dyons) can be written as  
\begin{equation} 
{\cal W}=\mu u(a_{D})+\sqrt{2} \tilde{M} a_{D} M   \label{sup} 
\end{equation} 
where $M$ is the monopole field. Minimization of the potential yields the 
monopole condensation, as a result, the standard Abrikosov-Nielsen-Olesen 
string appears with the tension proportional to the mass of the ${\cal N}=1$ 
chiral field $\mu$. 

We can expand the $u(a_{D})$ term in Eq. (\ref{sup}) as
\begin{displaymath}
\mu u(a_{D})=-\mu a_{D}+\eta a_{D}^{2}+\cdots \,,
\end{displaymath}
where the zero-th order approximation corresponds to the linear term in $a_{D}$
which preserves ${\cal N}=2$ supersymmetry ~\cite{c6}. Our task is to find 
the correction due to the quadratic term in $a_{D}$. In the Sec. 2 we 
investigate the supersymmetric electrodynamics (SQED) 
with only linear term in $a_{D}$.
The Abrikosov strings in this case are found to be $1/2$-BPS saturated. 
In the Sec. 3, we consider the additional quadratic term in the superpotential 
which destroys the BPS property. And the correction in the string tension in 
this case is calculated 
numerically to the first order of the perturbation parameter.

\section{$\frac{1}{2}$-BPS Saturated Abrikosov Strings}
\label{sec2} 
 
Consider ${\cal N}=2$ supersymmetric electrodynamics ~\cite{c7}. The ``photon'' 
$A_{\mu}$ is accompanied by its ${\cal N}=2$ superpartners(photinos)--two neutral 
Weyl spinors $\lambda$ and $\psi$, and a complex neutral scalar $a$. They form an 
irreducible ${\cal N}=2$ representation that can be decomposed as a sum of two 
${\cal N}=1$ representations: $a$ and $\psi$ are in a chiral representation, 
$\Phi$, while $A_{\mu}$ and $\lambda$ are in a vector representation, 
$W_{\alpha}$. Matter sector consists of two ${\cal N}=1$ chiral 
multiplets $M$ and $\tilde{M}$ with opposite electric charge. The 
renormalizable ${\cal N}=2$ invariant Lagrangian is described in an 
${\cal N}=1$ language by canonical kinetic terms and minimal gauge 
couplings for all the fields as well as a superpotential 
\begin{equation} 
{\cal W}=\sqrt{2}\Phi M \tilde{M}-\mu \Phi \,,   \label{lsup} 
\end{equation} 
here we replace the $a_{D}$ in Eq. (\ref{sup}) by $\Phi$ for simplicity.

The Lagrangian in component fields is given by 
\begin{eqnarray}
{\cal L}&\!\!=&\!\!-\frac{1}{4}F_{\mu\nu}F^{\mu\nu}+\partial^{\mu}\bar{a}
\partial_{\mu}a+i\lambda\sigma^{\mu}\partial_{\mu}\bar{\lambda}+i\psi\sigma^{\mu}
\partial_{\mu}\bar{\psi}  \nonumber   \\ 
 &\!\! &\!\!+D_{\mu}\bar{M}D^{\mu}M+D_{\mu}\bar{\tilde{M}}D^{\mu}\tilde{M}+i\psi_{M}
\sigma^{\mu}D_{\mu}\bar{\psi}_{M}+i\psi_{\tilde{M}}\sigma^{\mu}D_{\mu}
\bar{\psi}_{\tilde{M}}  \nonumber  \\ 
&\!\! &\!\!+\left(\sqrt{2}i\psi_{M}\lambda\bar{M}+h.c.\right)+
\left(-\sqrt{2}i\psi_{\tilde{M}}\lambda\bar{\tilde{M}}+h.c.\right)  \nonumber  \\ 
 &\!\! &\!\!+\left(\sqrt{2}a M F_{\tilde{M}}-\sqrt{2}a\psi_{M}\psi_{\tilde{M}}+
\sqrt{2}a F_{M}\tilde{M}-\sqrt{2}\psi\psi_{\tilde{M}}M \right.  \nonumber  \\ 
&\!\! &\!\!  \left. -\sqrt{2} \psi \psi_{M} \tilde{M}+\sqrt{2} F M 
\tilde{M}-\mu F+ h.c.\right)  \nonumber  \\ 
 &\!\! &\!\!+F\bar{F}+\frac{1}{2}D^{2}+D\left(\bar{M}M-\bar{\tilde{M}}
\tilde{M}\right)+F_{M}\bar{F}_{M}+F_{\tilde{M}}\bar{F}_{\tilde{M}}\,.   \label{lag} 
\end{eqnarray} 
Here we set the coupling constant $e=1$ for the convenience, which is not 
important in our later results. We stick to this convention in what follows. 
$F,F_{M},F_{\tilde{M}}$ and $D$ are the auxiliary fields which are given by 
\begin{eqnarray} 
\label{aux} 
F&\!\!=&\!\!\mu-\sqrt{2}\bar{M}\bar{\tilde{M}}\,,  \nonumber  \\ 
F_{M}&\!\!=&\!\!-\sqrt{2}\,\bar{a}\,\bar{\tilde{M}}\,,  \nonumber \\ 
F_{\tilde{M}}&\!\!=&\!\!-\sqrt{2}\,\bar{a}\,\bar{M}\,,   \nonumber \\ 
D&\!\!=&\!\!-\left(\bar{M}M-\bar{\tilde{M}}\tilde{M}\right)\,.   
\end{eqnarray}
 Here $M,\tilde{M}$ are the lowest components of the corresponding superfields, 
respectively, with the electric charges $\pm 1$, e.g. 
\begin{displaymath} 
D_{\mu}M=\partial_{\mu}M-iA_{\mu}M\,, \quad   D_{\mu}\tilde{M}=
\partial_{\mu}\tilde{M}+iA_{\mu}\tilde{M}\,. 
\end{displaymath} 
 
The scalar potential is minimized at  
\begin{equation} 
F=F_{M}=F_{\tilde{M}}=0\,, \quad   D=0\,, 
\end{equation} 
which occurs when 
\begin{equation} 
a=0\,, \quad  M\tilde{M}=\frac{\mu}{\sqrt{2}}\,,\quad  \mbox{and} 
\quad \left|M\right|=|\tilde{M}|\,. \nonumber 
\end{equation}

Then the supersymmetric transformations which preserve ${\cal N}=2$ 
supersymmetry are given by 
\begin{eqnarray} 
\delta a&\!\!=&\!\!\sqrt{2}\xi\lambda+\sqrt{2}\varepsilon\psi\,,  \nonumber  \\ 
\delta\psi&\!\!=&\!\!i\xi D-i\xi\sigma^{\mu\nu}F_{\mu\nu}+\sqrt{2}
\varepsilon F+\sqrt{2}i\sigma^{\mu}\bar{\varepsilon}\partial_{\mu}a\,,  \nonumber  \\ 
\delta F&\!\!=&\!\!-\sqrt{2}i\xi\sigma^{\mu}\partial_{\mu}
\bar{\lambda}+\sqrt{2}i\partial_{\mu}\psi\sigma^{\mu}\bar{\varepsilon}\,, \nonumber  \\ 
\delta A_{\mu}&\!\!=&\!\!-i\xi\sigma_{\mu}\bar{\psi}+i\psi\sigma_{\mu}\bar{\xi}-i\varepsilon
\sigma_{\mu}\bar{\lambda}+i\lambda\sigma_{\mu}\bar{\varepsilon}\,, \nonumber  \\ 
\delta\lambda &\!\!=&\!\!\sqrt{2}\xi\bar{F}+\sqrt{2}i
\sigma^{\mu}\bar{\xi}\partial_{\mu} a+iD\varepsilon+
i\sigma^{\mu\nu}F_{\mu\nu}\varepsilon\,,  \nonumber  \\ 
\delta D&\!\!=&\!\!\partial_{\mu}\psi\sigma^{\mu}\bar{\xi}+
\xi\sigma^{\mu}\partial_{\mu}\bar{\psi}+\partial_{\mu}\lambda\sigma^{\mu}
\bar{\varepsilon}+\varepsilon\sigma^{\mu}\partial_{\mu}\bar{\lambda}\,,  \nonumber  \\ 
\delta M&\!\!=&\!\!\sqrt{2}\varepsilon\psi_{M}+\sqrt{2}\bar{\xi}\bar{\psi}_{\tilde{M}}\,,  \label{trans}  \\ 
\delta \tilde{M}&\!\!=&\!\!\sqrt{2}\bar{\xi}\bar{\psi}_{M}+
\sqrt{2}\varepsilon\psi_{\tilde{M}}\,,  \nonumber  \\ 
\delta\psi_{M}&\!\!=&\!\!\sqrt{2}\varepsilon F_{M}+\sqrt{2}
i\sigma^{\mu}\bar{\varepsilon}D_{\mu} M+\sqrt{2}\sigma^{\mu}
\bar{\xi}D_{\mu}\bar{\tilde{M}}-2i\bar{a}\xi M\,,  \nonumber  \\ 
\delta\psi_{\tilde{M}}&\!\!=&\!\!\sqrt{2}\sigma^{\mu}
\bar{\xi}D_{\mu}\bar{M}+\sqrt{2}i\sigma^{\mu}\bar{\varepsilon}
D_{\mu}\tilde{M}+\sqrt{2}\varepsilon F_{\tilde{M}}-
2i\bar{a}\xi\tilde{M}\,, \nonumber \\ 
\delta F_{M}&\!\!=&\!\!\sqrt{2}iD_{\mu}\psi_{M}\sigma^{\mu}
\bar{\varepsilon}-2\bar{\xi}\bar{\lambda}\bar{\tilde{M}}-
2i\bar{a}\xi\psi_{M}\,, \nonumber  \\ 
\delta F_{\tilde{M}}&\!\!=&\!\!\sqrt{2}iD_{\mu}\psi_{\tilde{M}}
\sigma^{\mu}\bar{\varepsilon}-2\bar{\xi}\bar{\lambda}\bar{M}-
2i\bar{a}\xi\psi_{\tilde{M}}\,,  \nonumber 
\end{eqnarray} 
where the spinorial indices are suppressed. 
 
Without loss of generality, we can assume the Abrikosov string 
axis lies along the $z$ axis, while the string profile depends 
only on $x,y$. Then we obtain the saturation equations by requiring 
the fermionic fields transformations in Eq. (\ref{trans}) to vanish 
as follows 
\begin{eqnarray} 
F_{12}&\!\!=&\!\!\sqrt{2}\left(\sqrt{2}\bar{M}\bar{\tilde{M}}-\mu\right)\,, \nonumber \\ 
\left(D_{1}+iD_{2}\right)M&\!\!=&\!\!0\,,  \label{sequ}  \\ 
\left(D_{1}-iD_{2}\right)\tilde{M}&\!\!=&\!\!0\,,  \nonumber 
\end{eqnarray} 
with the constraint determining the parameter of the residual supersymmetry, 
\begin{equation} 
i\tau_{3}\xi =\varepsilon\,,  \label{con} 
\end{equation} 
 which reduces the number of supersymmetries from eight 
to four. The Abrikosov-Nielsen-Olesen string is $1/2$-BPS saturated. 
 
The Ansatz which goes through Eq. (\ref{sequ}) is 
\begin{eqnarray} 
M&\!\!=&\!\!\left(\frac{\mu}{\sqrt{2}}\right)^{\frac{1}{2}}e^{i\phi}\,\, f(r)\,,  \nonumber  \\ 
\tilde{M}&\!\!=&\!\!\left(\frac{\mu}{\sqrt{2}}\right)^{\frac{1}{2}}e^{-i\phi} f(r)\,,  \label{ansa}  \\ 
A_{\phi}&\!\!=&\!\!-2\,\frac{g(r)}{r}\,,  \nonumber
\end{eqnarray} 
with the boundary conditions

\begin{eqnarray}
f(0)&\!\!=&\!\!g(0)= 0\,,  \nonumber \\ 
\lim_{r \rightarrow \infty}f(r) &\!\!=&\!\!1\,, \nonumber  \\
\lim_{r\rightarrow\infty}g(r)&\!\!=&\!\!-\frac{1}{2}\,.  \nonumber 
\end{eqnarray}

The profile functions, $f(r)$ and $g(r)$ satisfy the first-order 
differential equations 
\begin{eqnarray} 
f^{\prime}&\!\!=&\!\!\frac{f}{r}(1+2g)\,,  \nonumber  \\ 
g^{\prime}&\!\!=&\!\!\frac{1}{2}r(1-f^2)\,,  \label{diff} 
\end{eqnarray} 
where the prime denotes differentiation over $r$. 
 
One can calculate the string tension as follows 
\begin{eqnarray} 
{\cal T}&\!\!=&\!\!\int d^2 x \left\{\frac{1}{2}F_{12}^{2}+D_{1}\bar{M}D_{1}M+D_{2}\bar{M}D_{2}M+
D_{1}\bar{\tilde{M}}D_{1}\tilde{M}+D_{2}\bar{\tilde{M}}D_{2}\tilde{M} \right. \nonumber  \\ 
 &\!\! &\!\! \left. +\left(\mu-\sqrt{2}\bar{M}\bar{\tilde{M}}\right)\left(\mu-\sqrt{2} 
 M\tilde{M}\right)+\frac{1}{2}\left(\bar{M}M-\bar{\tilde{M}}\tilde{M}\right)^{2}\right\} \nonumber \\ 
 &\!\!=&\!\!\int d^2 x\left\{\left|\frac{1}{\sqrt{2}}F_{12}-\left(\sqrt{2}\bar{M}
 \bar{\tilde{M}}-\mu\right)\right|^{2}+\left(D_{1}+iD_{2}\right)\bar{M}\left(D_{1}+iD_{2}\right)M \right.  \nonumber  \\ 
 &\!\! &\!\!+\left(D_{1}-iD_{2}\right)\bar{\tilde{M}}\left(D_{1}-iD_{2}\right)\tilde{M}+
 \frac{1}{2}\left(\bar{M}M-\bar{\tilde{M}}\tilde{M}\right)^{2}-\sqrt{2}\mu
 F_{12}  \label{ten}  \\ 
 &\!\! &\!\!- \left. i\left[\partial_{1}\left(\bar{M}D_{2}M\right)-\partial_{2}\left(\bar{M}D_{1}
 M\right)\right]+i\left[\partial_{1}\left(\bar{\tilde{M}}D_{2}\tilde{M}\right)-\partial_{2}
 \left(\bar{\tilde{M}}D_{1}\tilde{M}\right)\right]  \right\} \,.  \nonumber 
\end{eqnarray} 
Applying Eq. (\ref{sequ}) and neglecting the total derivative terms, we get
\begin{equation} 
{\cal T}=-\sqrt{2}\mu\int d^2 x F_{12}\,.   \label{sten} 
\end{equation}

\section{Small Perturbation in Superpotential} 
\label{sec3} 
 
We can add a small perturbation in the superpotential Eq. (\ref{lsup}) 
\begin{equation} 
{\cal W}=\sqrt{2}\Phi M \tilde{M}-\mu\Phi+\eta\Phi^{2}\,,  \label{qsup} 
\end{equation} 
where $\eta$ is a real small perturbation parameter. 
 
Then we can go over the analysis in Sec. 2 in the similar way. 
But the small perturbation will break ${\cal N}=2$ supersymmetry and 
the resultant Abrikosov string is no longer BPS saturated which can 
be seen clearly in the string tension  
\begin{eqnarray}
{\cal T}&\!\!=&\!\!\int d^2 x \left\{ \frac{1}{2}F_{12}^{2}+D_{1}\bar{M}D_{1}M+D_{2}\bar{M}D_{2}M+
D_{1}\bar{\tilde{M}}D_{1}\tilde{M}+D_{2}\bar{\tilde{M}}D_{2}\tilde{M} \right. \nonumber  \\ 
 &\!\! &\!\!+\left(\mu-\sqrt{2}\bar{M}\bar{\tilde{M}}-2\eta\bar{a}\right)\left(\mu-\sqrt{2} M\tilde{M}-
 2\eta a\right)+\frac{1}{2}\left(\bar{M}M-\bar{\tilde{M}}\tilde{M}\right)^{2} \nonumber \\ 
 &\!\! &\!\!+ \left. \partial_{1}\bar{a}\partial_{1} a+\partial_{2}\bar{a}\partial_{2} a+2a\bar{a}M\bar{M}+
 2a\bar{a}\tilde{M}\bar{\tilde{M}}\right\} \nonumber  \\ 
 &\!\!=&\!\!\int d^2 x\left\{\left|\frac{1}{\sqrt{2}}F_{12}-\left(\sqrt{2}
 \bar{M}\bar{\tilde{M}}-\mu\right)\right|^{2}+
 \left(D_{1}+iD_{2}\right)\bar{M}\left(D_{1}+iD_{2}\right)M \right.  \nonumber  \\ 
 &\!\! &\!\!+\left(D_{1}-iD_{2}\right)\bar{\tilde{M}}\left(D_{1}-iD_{2}\right)\tilde{M}+\frac{1}{2}\left(
 \bar{M}M-\bar{\tilde{M}}\tilde{M}\right)^{2}-\sqrt{2}\mu F_{12}  \nonumber  \\ 
 &\!\! &\!\!-i\left[\partial_{1}\left(\bar{M}D_{2}M\right)-\partial_{2}\left(\bar{M}D_{1}M\right)\right]+i\left[
 \partial_{1}\left(\bar{\tilde{M}}D_{2}\tilde{M}\right)-\partial_{2}
 \left(\bar{\tilde{M}}D_{1}\tilde{M}\right)\right]  \nonumber  \\ 
  &\!\! &\!\!+ \left. \partial_{1}\left(\bar{a}\partial_{1}a\right)+
 \partial_{2}\left(\bar{a}
 \partial_{2}a\right)-2\eta a\left(\mu-\sqrt{2}\bar{M}\bar{\tilde{M}}\right) \right\}\,,  \label{qten} 
\end{eqnarray}
where we have used the equation of motion for field $a$ 
\begin{equation} 
-\partial_{1}^{2}a-\partial_{2}^{2}a=2\eta\left(\mu-\sqrt{2}M\tilde{M}-2\eta 
a\right)-2a M\bar{M}-2a\tilde{M}\bar{\tilde{M}}\,.  \label{aequ} 
\end{equation} 
 
To the first order of $\eta$, we can still use the Ansatz (\ref{ansa}) 
for the fields $M,\tilde{M}$ and $A_{\mu}$. After applying Ansatz 
(\ref{ansa}), Eq. (\ref{aequ}) becomes to the first order in $\eta$ 
\begin{equation} 
\partial_{1}^{2}a+\partial_{2}^{2}a=-2\eta\mu\left(1-f^{2}\right)+
2\sqrt{2}\mu a f^{2}\,.  \label{aequ2} 
\end{equation}

Then the string tension turns out to be  
\begin{equation} 
{\cal T}=\int d^{2} x \left\{\left(-\sqrt{2}\mu F_{12}\right)-2\eta 
a\left(\mu-\sqrt{2}\bar{M}\bar{\tilde{M}}\right)\right\}\,.  \label{qten} 
\end{equation} 
 
Then from Eq. (\ref{sten}),(\ref{qten}), we can find the difference of the 
string tensions between non-BPS and BPS saturated situation to be 
\begin{eqnarray} 
\Delta {\cal T}&\!\!=&\!\!\int d^{2} x \left\{-2\eta a\left(
\mu-\sqrt{2}\bar{M}\bar{\tilde{M}}\right)\right\}  \nonumber  \\ 
 &\!\!=&\!\!-2\eta \mu\int d^{2} x\, a \left(1-f^{2}\right)\,,  \label{dten} 
\end{eqnarray} 
where we have used the Ansatz (\ref{ansa}). 
 
To see Eq. (\ref{dten}) more clearly, one can switch to dimensionless quantities 
\begin{eqnarray} 
x&\!\!\rightarrow&\!\!\frac{1}{(\sqrt{2}\mu)^{\frac{1}{2}}} x\,,  \nonumber \\ 
y&\!\!\rightarrow&\!\!\frac{1}{(\sqrt{2}\mu)^{\frac{1}{2}}} y\,, \nonumber  \\ 
a&\!\!\rightarrow&\!\!\eta a\,.  \label{dimen} 
\end{eqnarray} 
Then we get 
\begin{equation} 
\Delta{\cal T}=-\sqrt{2}\eta^{2}\int a\left(1-f^{2}\right) d^{2} x\,,  \label{tendiff} 
\end{equation} 
where $a,f$, and $x$ here are dimensionless.
 
We can solve Eq. (\ref{diff}),(\ref{aequ2}), and calculate $\Delta 
{\cal T}$ in Eq. (\ref{tendiff}). The result is 
\begin{equation} 
\Delta {\cal T}=-2\sqrt{2}\pi\eta^{2}\, 0.68 <\, 0\,. 
\end{equation}

\section{Conclusions} 
\label{sec4} 
 
We investigated the Abrikosov-Nielsen-Olesen string solution in ${\cal N}=2$ 
supersymmetric electrodynamics with some ${\cal N}=2$-preserving superpotential. 
The string solution is due to the superpotential rather than due to the 
Fayet-Iliopoulos term. The Abrikosov string was found to be $1/2$-BPS 
saturated which follows directly from the ${\cal N}=2$ supersymmetric 
transformations. After the ${\cal N}=2$ supersymmetry is broken to ${\cal N}=1$ 
by the perturbation in the superpotential, the Abrikosov string is no longer 
BPS saturated. And the string tension in this case was found to be less than that of the BPS case. 
 
\section{Acknowledgments} 
I am grateful to M. Shifman for suggesting the problem to me and for numerous
useful discussions. I would also like to thank A. Vainshtein and A. Yung for
useful discussions.
This work was supported in part by DOE under the grant number DE-FG02-94ER408.

\end{document}